# Broadband, Temperature Tolerant and Passively Biased Resonantly Enhanced Mach-Zehnder Modulators

S. Romero-García, A. Moscoso-Mártir, J. Nojic, S. Sharif-Azadeh, J. Müller, B. Shen, F. Merget, J. Witzens

Institute of Integrated Photonics
RWTH Aachen University
Aachen, Germany

*Abstract*—We describe a resonantly enhanced Mach-Zehnder modulator (MZM) that can be operated over a wide temperature range of 55°C without being actively biased, while providing a significant resonant enhancement of 6.8 at the nominal wavelength / temperature compared to a linear MZM driven with a distributed driver. More importantly, it enables a ~20X improvement in power consumption compared to a 50 Ω matched linear travelling wave modulator with comparable phase shifter technology, drive voltage and output optical modulation amplitude. Passive biasing of the Mach-Zehnder interferometer is further implemented by replacing a splitter element in the MZM with a novel device combining splitting and fiber coupling functionalities in a single, multi-modal structure, that converts permanent fiber placement into a phase correction. Both concepts are combined in a single modulator device, removing the need for any type of active control in a wide temperature operation range.

*Keywords—Electro-optic modulators; Datacom; Resonant ring modulators; Electro-optic transceivers.*

## I. INTRODUCTION

Reducing the power consumption of electro-optic transceivers in data centers has become a key objective on par with the reduction of manufacturing cost: Since interconnects have been estimated to consume 23% of the total datacenter energy intake [1], their power consumption is a burden on both operating expenses (OpEx) and the required cooling infrastructure.

Travelling wave (TW) modulators can make up a substantial portion of an electro-optic transmitter's power consumption, which has motivated the investigation of alternative modulator and driver topologies enabling a significant reduction of the required power. Distributed drivers are one such alternative [2] that has already found its way into industrial practice [3]. This allows reducing the power consumption associated to dissipation of the high-speed signal delivered to the modulator (the RF power consumption) by a factor 6 to 8 [4]. Another approach consists in shrinking the size of the modulator to a fraction of the radio-frequency (RF) wavelength, to enable driving it as an electrical lumped element with a nearby low output impedance driver. In order to overcome the reduction of modulation efficiency resulting from shrinking of the phase shifter length, resonantly enhanced devices such as resonant ring modulators (RRMs) have attracted considerable attention in silicon photonics [5]. Discussions in the literature typically focus on the resonant enhancement itself, i.e., the increase in modulation efficiency as compared to a linear, non-resonant modulator with an equally long embedded phase shifter length. Equivalently, this essentially also corresponds to the power consumption reduction as compared to a linear modulator with a phase shifter length chosen to result in the same modulation contrast, *assuming it can be driven as a lumped element*. Another important improvement, however, arises from the much-reduced dimensions of resonantly enhanced modulators, which allow them to be driven as a lumped element in the first place. While longer linear devices can be driven with a similar RF power consumption, as associated to the signal delivered to the modulator, as a lumped element modulator by means of a distributed driver, a single lumped element load allows achieving this power consumption improvement without incurring the additional integration or packaging complexity associated to distributed drivers, i.e., the monolithic integration of a distributed driver via CMOS photonic process integration [3] or the more conventional route of co-packaging of the distributed driver via flip-chipping and a micro-bump array. A lumped element modulator, on the other hand, can be simply integrated with its driver via a pair of wire-bonds, provided the capacitive load of the modulator remains sufficiently small to tolerate the parasitic inductance of the wire-bonds [6]. While the RF power consumption inside the modulator itself only constitutes a portion of the power drawn by the modulator driver, at least the output stage of the driver needs to be sized according to the load, so that it remains an important quantity.

In addition to the RF power consumption, additional power is associated to the stabilization of the modulators against temperature changes and process bias: Silicon possesses a relatively high thermo-optic coefficient of 1.87e-4 RIU/K at room temperature [7] and the effective index of silicon/$SiO_2$ single mode waveguides is particularly prone to roughness induced variations due to the high index contrast, so that index variations in interferometric and resonant devices have to be compensated for. This is a tractable problem in linear modulators, that are nominally balanced so that variations induced by temperature swings are relatively modest and can be straightforwardly compensated for. Self-referenced resonant devices on the other hand are very sensitive to temperature changes and process biases: The typical resonance wavelength non-uniformity across dies due to fabrication and layer thickness is on the order of ±1 nm [8]. Although advanced schemes have been developed to reduce phase tuning requirements, e.g. in multi-channel WDM transceivers [9], in a typical system configuration wavelength tuning by a full free spectral range

(FSR) is required at start-up, followed by additional tuning during operation to compensate for environmental temperature swings. While thermally isolating resonant devices has resulted in very substantial improvements of the power efficiency of thermal tuners [10], this only partially alleviates the problem, as on the one hand process complexity is increased and on the other hand limitations to the maximum allowable device temperature as constrained by device reliability remain. Finally, athermal resonant devices have been realized by means of cladding materials, typically polymers, with opposite thermooptic coefficients [11]. Here too, process complexity is increased and long term reliability associated to the introduction of polymers needs to be ascertained.

For the aforementioned reasons, a lumped element, resonantly enhanced device that does not require dynamic tuning and that can be fabricated in a standard silicon photonics fabrication line would be highly desirable. Here, we combine two techniques to achieve this objective: We have previously shown that multi-mode fiber/laser-to-chip couplers initially developed to relax alignment tolerances [12],[13] can also be used to passively and permanently bias a Mach-Zehnder interferometer [14]. Moreover, we have adapted a device architecture in which collectively driven RRMs are utilized as phase shifters [15] to obtain optically broadband operation while maintaining a sizeable resonant enhancement [16] and high-speed, lumped element operation [6]. In the following, we will first review results pertaining to these two schemes and continue with a description of the combined device.

## II. PASSIVE BIASING OF MZMs BY MEANS OF MULTIMODE COUPLERS

In [12],[13] we showed laser-to-chip edge couplers and fiber-to-chip grating couplers that have in common their ability to relax required alignment tolerances in one direction (transverse relative to the main optical axis of the devices) and the fact that they both have two single mode output waveguides. The couplers were architectured in such a way that the power coupled to both single mode output waveguides is substantially equal, but that the phase of the light coupled to either waveguide varies as a function of the placement of the input fiber or input laser. Since we sought to improve alignment tolerances without reducing the peak insertion efficiency, this degree of freedom was required so as to not violate the reciprocity theorem. The devices are multi-mode in nature and essentially combine a coupler and a Y-junction without forcing the light through a unique single mode optical path in-between. The operation principle of the devices can be found in the references [12],[13]. In a parallel single mode optics (PSM) transceiver, this varying phase at the output of the coupler is irrelevant, since the light is never recombined: One seeks to guarantee a minimum power in any of the downstream parallel communication channels (obtained by further splitting of the light).

In [14] we repurposed these devices in order to use the phase shift at their output as a feature: We implemented a Mach-Zehnder interferometer (MZI), in which the input splitter was replaced by one such multi-mode grating coupler (MMGC). By moving the position of the input fiber across the MMGC, the relative phase at the two output waveguides is varied until the quadrature point (biasing at -3 dB MZI output power) is reached, after which the fiber is affixed in place by means of a UV-curable epoxy.

Adequacy of this technique is contingent on a few elements: The fiber position can be assumed to be reliably fixed in place since packaging based on UV curable epoxy is an established technology. Moreover, the couplers themselves can be designed to be sufficiently temperature tolerant so that their functionalities, 3 dB splitting and a fixed phase offset at the output, is maintained over a wide temperature range. However, one also has to assume that the mismatch of the interferometer is of a sufficiently low order for the static correction of a fixed phase term to rebalance it over all relevant temperatures and wavelengths (which is easy to verify or to invalidate for a given interferometer). Lastly, one has to assume that the phase mismatch occurring in the two arms of the interferometer does not drift over time due to additional effects, such as for example the diffusion of fixed charges in surrounding dielectrics, that can not only cause losses, but also effective index changes [17]. This is the subtlest assumption to verify as it relates to long term reliability and remains an open question. Lastly, this scheme requires that at least one fiber per interferometer can be freely positioned, which would for example prevent packaging of a modulator array with a single fiber array. This may be the primary limitation of this approach. An important strength lies however in the fact that the phase correcting mechanism itself should be very robust and compatible with long term reliability.

The concept of the alignment tolerant grating coupler is shown in Fig. 1. As can be seen in the bottom panel, the power coupled in either waveguide is equal and remains within 1 dB of its nominal value (centered fiber) within a ±7.2 μm misalignment range in the transverse/lateral direction (red arrow in the micrograph), a tolerance almost 3X larger than a conventional grating coupler. In the longitudinal direction, the device behaves like a conventional grating coupler, i.e., the alignment tolerance is not enhanced. The total coupled power is only slightly worse to that of the conventional grating coupler, with the slight excess losses attributed in particular to the built-in Y-junction (highlighted with a black rectangle in the micrograph). Importantly, as the fiber is moved transversely across the grating coupler, the relative phase of the light coupled into the two output waveguides is varied, which is utilized in the following (in which case the alignment tolerance is again lost, as now placement is constrained by other considerations than coupling efficiency).

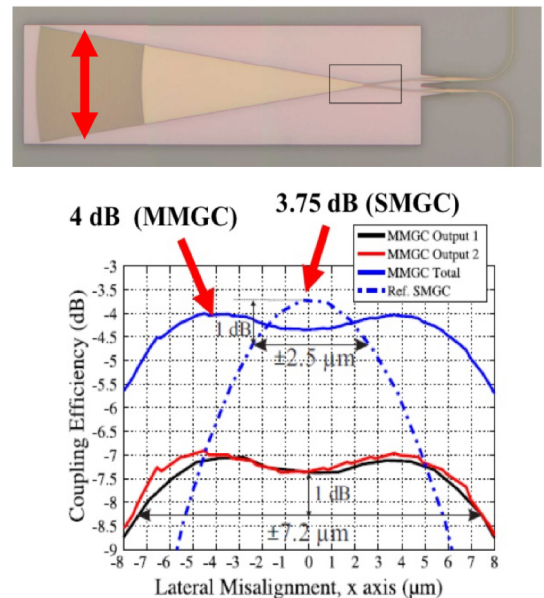

Fig. 1. (top) Micrograph of an alignment tolerant grating coupler (MMGC) as reported in [13] and (bottom) corresponding coupling efficiency into either of the two output waveguides (black and red curves) as well as the cumulative coupling efficiency (solid blue curve) as compared to a regular single mode grating coupler (SMGC) fabricated in the same technology (dash dotted blue curve). The transverse misalignment, as shown by the red arrow in the micrograph, is indicated on the x-axis.

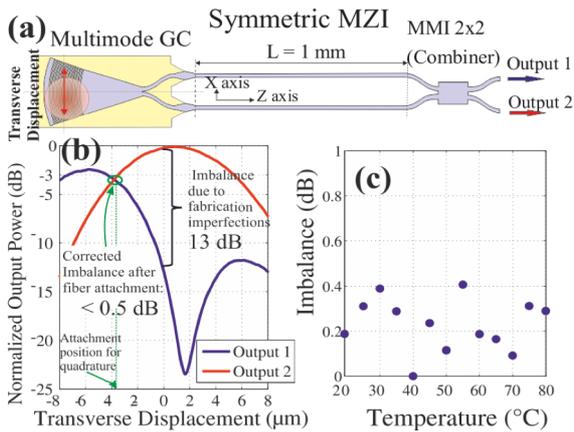

Fig. 2. (a) MZI with multimode grating coupler as input splitter. (b) Measured MZI output power levels (relative to the maximum transmission) as a function of the input fiber position. (c) Imbalance between the two MZI output ports as a function of temperature after permanent fiber attachment.

The concept of the entire MZI is illustrated in Fig. 2(a), experimental results are shown in Fig. 2(b). As the fiber is moved along the transverse direction above the grating coupler, the phase in the two branches of the MZI is varied until the power at the output of the MZI, after an additional 2 by 2 multimode interferometer (MMI), is exactly balanced. The fiber is then epoxied in place and the temperature dependent characteristics of the interferometer are measured.

As seen in figure 2(b), the initial imbalance of the device, as measured with the optical fiber centered on the optical axis of the grating coupler, was 13 dB as a consequence of fabrication induced mismatch between the two 1 mm long arms of the interferometer. After displacement and permanent attachment of the fiber this imbalance was maintained below 0.4 dB even while cycling the device temperature between 20°C and 80°C. This shows that in this case the imbalance order of the MZI was sufficiently low for the MZI to remain balanced over the entire temperature range and confirms that the temperature tolerance of the grating coupler, also in terms of its output phases, is sufficient for this purpose. For details on the devices, please refer to references [12]-[14].

A number of other techniques have been shown to modify the bias point of passive interferometers or resonant devices. These comprise local oxidation of the waveguide [18], optically induced modification of the refractive index of a thin chalcogenide film surrounding the waveguide by means of UV radiation [19], thermally induced refractive index change in the cladding [20], or thermally/current induced dopant diffusion [21]. While the method shown in [20] does not yet feature long term stability and the method shown in [21] was self-limiting and did not yet yield large phase shifts, they are interesting in that they target standard silicon photonics process flows without further modifications. The method shown in [18] in particular can be expected to feature long term stability. In all these cases, long term stability of the trimming mechanism itself and of other phase shifts within the device is a primary concern for the practicability of the trimming mechanism.

### III. TEMPERATURE TOLERANT, RESONANTLY ENHANCED MZMs

The phase shifters in the MZM are implemented in the form of collectively driven, highly overcoupled RRMs loaded on a common bus waveguide [16]. In the highly overcoupled regime, when the optical carrier is tuned on resonance, the transfer function of a single ring resonator can be given as [22]

$$A_{Out} = A_{In} e^{i\left(\beta - \frac{\alpha}{2}\right)\frac{2F}{\pi}L}$$

where $A_{In}$, $A_{Out}$ are the field amplitudes in the bus waveguide, respectively before and after the ring, $\beta$, $\alpha$ are respectively the average wave number and linear losses inside the ring, $F$ is the finesse given as the ratio of the FSR to the full width at half maximum (FWHM) and $L$ is the circumference of the ring. Thus, it can be seen that the ring behaves as a linear waveguide whose length $L$ has been demultiplied by a factor $2F/\pi$, hence the resonant enhancement factor applied to an embedded phase shifter section. Importantly, in the absence of excess losses due to e.g. waveguide bends or excess losses in the ring to waveguide coupling junction, the $V_\pi L \cdot \alpha$ figure of merit of the phase shifter remains unchanged, wherein $V_\pi L$ is the drive voltage required to achieve a phase shift of $\pi$ in a phase shifter of length $L$.

In order to obtain a high optical bandwidth, an increased FWHM and thus a reduced resonator quality (Q-)factor are required. This also reduces the resonant enhancement factor $2F/\pi$ (assuming the embedded phase shifter to cover the entire circumference) since the finesse scales as the opposite of the FWHM. This can be compensated by increasing the FSR, i.e., reducing the circumference $L$ of the ring (this ought to make sense intuitively, as the capacitance of the embedded phase shifter, and thus its power consumption, scales with its length). While this is conceptually simple, its reduction to practice is quite tricky: As the circumference shrinks, the sensitivity of the device to increased coupling losses rises (since the excess losses are distributed over the waveguide length to convert them into effective waveguide losses). In addition, bending losses go up, also contributing to increase $V_\pi L \alpha$. Finally, it becomes increasingly difficult to obtain sufficiently high coupling strengths over shrinking distances in order to maintain the high overcoupling regime required to spoil the resonator Q-factor without incurring excess losses. The design problem can thus be formulated as shrinking the device, while minimizing excess losses to maintain the same $V_\pi L\alpha$ and sufficiently overcoupling the ring.

A diagram and a scanning electron microscope (SEM) image of the fabricated waveguides (prior to back-end-of-line fabrication) can be seen in Figs. 3(a) and 3(b). The overlay with implant regions and the position of the electrodes can be seen in Fig. 3(c) and a micrograph of the complete RRM in Fig. 3(d). Several tricks were applied to achieve the design goals: Away from the coupler region, the waveguide inside of the ring is widened in order to reduce bending losses. This, however, results in too small a coupling parameter, so that the waveguide has to be narrowed again in the coupling region. In order to suppress resulting bending losses, the silicon is fully etched on the opposite side of the bus waveguide, which can be done while maintaining electrical connectivity of the embedded PIN diode. This does result, however, in the waveguide region inside the coupling region to have an increased series resistance to the electrodes and thus a reduced RC limited cutoff frequency. In order to prevent this from impacting the electro-optic $S_{21}$ of the device, as well as to prevent insertion losses associated to this inefficiently modulated region, it is left unimplanted (Fig. 3(c)). The device reported in [16] was fabricated in the standard silicon photonics process of IME A*STAR with a 220 nm silicon device layer thickness and a shallow etch depth of 130 nm (90 nm slabs). The outer radius of the resonators was chosen as R = 5 µm, the waveguide width varied between $W_1$ = 425 nm (in the coupling section) and $W_2$ = 1.575 µm. The bus waveguide width in the coupling regions was chosen as 370 nm and the bus waveguide to ring resonator gap as 200 nm to allow fabrication with 248 nm DUV lithography. p- and n-regions were doped with concentrations of respectively 2.5e18 cm$^{-3}$ and 2e18 cm$^{-3}$ and were spaced by an intrinsic region of nominally 20 nm (same phase shifter design as the third category of devices in [23]) and cover ¾ of the ring's circumference.

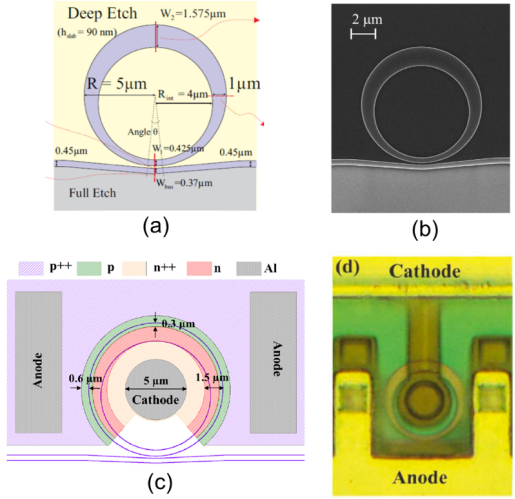

Fig. 3. (a) Diagram of the waveguide in a single RRM layout from [16], (b) scanning electron micrograph after waveguide fabrication, (c) mask layout showing overlay with the PIN junction as well as electrode placement, and (d) micrograph of the complete RRM.

As mentioned above, excess losses occurring at the waveguide to ring junction are critical in order to avoid spoiling the $V_\pi L \alpha$ of the phase shifter. In order to reduce these, a wiggle was introduced in the bus waveguide shape in order to smoothly taper the gap size [24]. As shown in Fig. 3(a), the bus waveguide follows a circular shape with the same center as the ring over a small angle $\theta$, outside of which it continues on a straight line prior to being bent back to the next resonator. Figs. 4(a) and 4(b) show simulated excess round trip losses and power coupling strengths for different resonator designs and $\theta$ values, including for the nominal design described above with $W_1$ = 425 nm, $W_2$ = 1.575 µm, R = 5 µm and $\theta$ = 5°. Figs. 4(c) and 4(d) show corresponding experimental data extracted from fitted resonances, also for $W_1$ = 425 nm, $W_2$ = 1.575 µm and R = 5 µm.

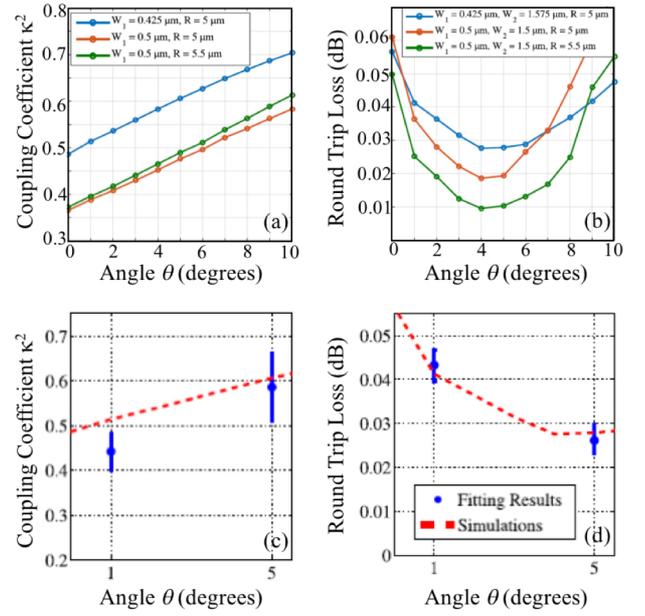

Fig. 4. (a) Simulated bus to ring waveguide power coupling coefficient and (b) simulated round trip excess losses as a function of $\theta$. (c) and (d) show the corresponding experimental data overlaid with simulation data for the nominal ring design corresponding to $W_1$ = 425 nm, $W_2$ = 1.575 µm, and R = 5 µm.

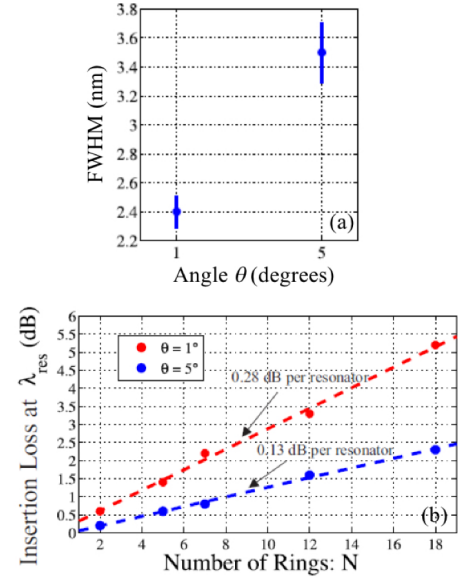

Fig. 5. (a) FWHM of the measured ring resonators ($W_1$ = 425 nm, $W_2$ = 1.575 µm, and R = 5 µm as in Fig. 4) as a function of $\theta$ and (b) insertion losses as a function of the number of resonators $N$.

The resulting FWHM and excess modulator insertion losses, with the optical carrier on resonance, are shown in Fig. 5(a) and 5(b), wherein the resulting resonator excess insertion losses already contain the resonant enhancement factor (these correspond to the losses incurred at the bottom of the Lorentzian resonator transfer function multiplied by the number of resonators).

The complete modulator was implemented in a subsequent fabrication run with 5 collectively driven resonant ring modulators loaded on each arm, resulting in a cumulative PIN junction length of 110 μm on each interferometer arm, total on-resonance modulator insertion losses of 5.7 dB, only 1.2 dB of which were attributed to excess losses (the rest corresponding to the resonantly enhanced losses of the embedded phase shifters), total capacitance and resistance per modulator arm of respectively 72 fF and 48 Ω, and a resonator FWHM of ~2 nm. The reduction of the FWHM as compared to the 3.5 nm measured in the previous run is attributed to a reduction of the coupling coefficient $\kappa^2$ from ~0.6 to ~0.4 due to process biases. The actual optical bandwidth of the modulator, defined as the optical carrier wavelength range in which the modulation efficiency remains within a factor 2 of its maximum, was measured as 3.9 nm. The $V_\pi L$ for an on-resonance optical carrier was measured as 0.19 V·cm (as normalized relative to the cumulative PIN junction length), 6.8 times lower than a linear phase shifter with the same PIN junction design. This enhancement was actually slightly below the actual resonant enhancement of 8, due to a slight reduction in phase shifter efficiency associated to the reduced optical field confinement in the wider sections of the ring (compared to the reference modulator with a 400 nm waveguide width). The modulator could be operated with a cutoff frequency of 23.5 GHz in a 50 Ω environment, as limited by the RC time constant associated to the capacitance of the 5 parallel rings and the sum of the 48 Ω total phase shifter resistance with the 50 Ω of the driving circuitry. With a low output voltage lumped element driver one would expect the RC time constant to be significantly improved, up to the 46 GHz intrinsic phase shifter cutoff frequency. We showed co-operability of the device with a wire bonded 25 Gbps low output impedance lumped element driver, but increased data rates could not yet be shown due to the speed limitations of the utilized driver. With 2 $V_{pp}$ and 4 $V_{pp}$ drive signals, the modulator yielded modulation penalties associated to the finite drive voltages (as measured as a reduction in the optical modulation amplitude) of respectively -4.8 dB and -2.2 dB for an on-resonance optical carrier.

The factor 6.8 reported above essentially corresponds to the reduction in RF power consumption associated to the signal delivered to the modulator, as compared to a linear modulator with identical PIN junction design (embedded in high confinement, 400 nm wide waveguides) and sized to have identical modulation penalty. The insertion losses of the resonantly enhanced modulator, 1.2 dB higher than the equivalent linear modulator, however also have to be factored in to yield a fair comparison. Increasing the drive voltage from 2 $V_{pp}$ to 2.7 $V_{pp}$ would for example compensate for these additional insertion losses, provided the driver technology is able to support such a voltage increase, resulting in a reduced effective power enhancement slightly below 4. From this perspective, the improvement of the device seems very modest, particularly in view of the remaining thermal sensitivity. This would however be missing the main improvement, as the comparison above is really between the resonantly enhanced modulator and an equivalent linear modular, *assuming it can be driven as a lumped element*, which, with a cumulative length of > 500 μm would not be the case at RF frequencies above a few GHz. Thus, in order for this comparison to hold, the linear modulator would need to be driven by a distributed driver. In other words, the resonantly enhanced device provides the benefits of lumped element driving without requiring a distributed driver. Compared to a TW wave device, the ~8.1 mW required by the resonantly enhanced device at 25 Gbps assuming a 3 $V_{pp}$ drive voltage is a very substantial improvement (20X compared to a dual drive, 50 Ω, linear TW MZM driven with a 2 $V_{pp}$ signal taking into account the reduction of the drive signal due to lower insertion losses).

The improved temperature tolerance should also be discussed: In [16] we showed that the optical modulation amplitude (OMA) at the output of the device remained within a factor two of its maximum in a 3.8 nm wavelength range at fixed temperature, or, equivalently, in a 55°C temperature range if the wavelength of the laser remains fixed. While this represents more than a factor 10 increase in wavelength / temperature tolerance compared to a typical high-speed RRM with a Q-factor of ~5000 [23], the question remains whether this is sufficient to operate a transceiver completely without tuning of the rings. The typical wavelength repeatability of off-the-shelf laser diodes is on the order of ±1 nm to ±2 nm. Since the repeatability of silicon photonics resonant devices across different dies and wafers is also on the order of ±1 nm, the initial wavelength tolerance can be seen to be largely eaten up by these sources of variabilities. While the remaining wavelength / temperature tolerance of below 1 nm / 14°C is too low on its own, a few elements may help: For one, since the laser and the modulator are operated in the same transmitter, temperatures can be assumed to largely co-vary in both devices, so that wavelength drifts will also be comparable given the similar thermo-optic coefficients of Si and InGaAsP [25]. Moreover, one may envision implementing the laser as an external cavity laser with the resonance wavelength determined by the silicon photonics chip [26], in which case variation of the wavelength as caused e.g. by varying Si film thicknesses might be applied to both devices, in particular if the tunable reflector closing the laser cavity and the modulator are laid out close to each other. Nonetheless, increasing the optical bandwidth of the device appears desirable, not only to increase the robustness of the system, but also so as to not incur the full additional 3 dB modulation penalty associated to operating the device at the edges of its optical passband.

Fortunately, this can be improved by increasing the number of rings loaded on each of the modulator arms, with the only penalty being, at moderate increase, a corresponding increase in power consumption, which, starting from ~8 mW is quite acceptable. Practical difficulties associated to increasing the number of rings are expected to be related to driving the increased capacitance, for example increasing the difficulty of designing a fitting low output impedance driver due to increase current sourcing requirements or worsening the detrimental effect of parasitic inductances associated to packaging.

Figure 6 shows the modeled optical OMA at the output of the temperature tolerant MZM described above, under the assumption of a 2 $V_{pp}$ drive voltage and further assuming that the number of collectively driven RRMs is varied between 3 and

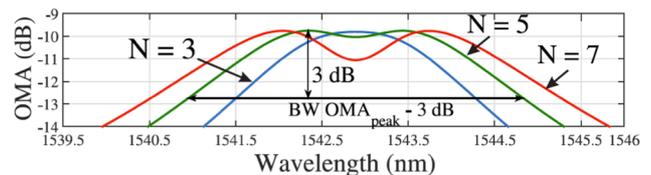

Fig. 6. OMA as a function of the number of collectively driven RRMs per MZM branch, assuming a drive voltage of 2 $V_{pp}$.

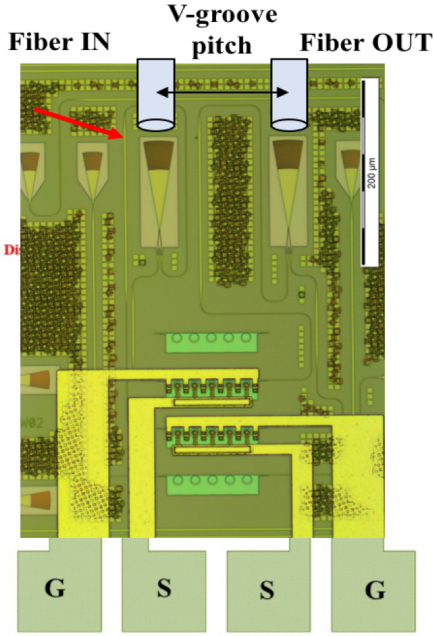

Fig. 7. Micrograph of the passively biased, temperature tolerant resonantly enhanced Mach-Zehnder modulator. Two fibers with a fixed pitch constrained by a fiber array are positioned on top of multi-mode grating couplers also acting as splitter / combiner for the interferometer. One of the interferometer branches is routed around the first MMGC (red arrow) in order to maintain the phase sensitivity of the device [13].

7 per MZM arm (5 being the nominal value for the experimental results reported above). Importantly, as mentioned above, as the number of RRMs is increased, the optical bandwidth of the device widens. At the same time, the OMA at the center frequency drops, which is not problematic at first if it remains above the minimum specified in the definition of the passband.

This can be understood with the following considerations: The OMA of a linear modulator with phase shifters of length $L$ driven in dual drive configuration with a drive voltage swing $V_d$ can be simply calculated as

$$OMA[dB] = 10 log_{10}(sin(V_d \cdot L \cdot \pi / V_\pi L)) \quad (2)$$
$$- 10 \cdot \alpha \cdot L / log(10)$$

where $\alpha$ is the linear loss term associated to the phase shifters. Even when ignoring the nonlinearity introduced by the sine function, one can straightforwardly see that this results in an optimum modulator length due to the different scaling of $log(L)$ and $L$. This is also what drives the trends in Fig. 6: As the number of rings is increased, and thus also the total effective phase shifter length at resonance $2NFL_{RRM}/\pi$, where $L_{RRM}$ is the circumference of a single RRM, the phase shifter length grows beyond it optimum for an on-resonance carrier frequency. For highly detuned carrier frequencies on the other hand the effective phase shifter length is lower, as the resonant enhancement is significantly reduced. Thus, for these frequencies the effective phase shifter length is still below optimum and the phase shifter efficiency continues to grow with the number of rings. The OMA is maximized for two intermediate frequencies at which the effective phase shifter length is exactly optimum, with these two frequencies moving away from the center frequency as the number of rings is increased.

## IV. COMBINED DEVICE

As a final step, we combined the two concepts described in the following sections in a single device. A micrograph of the combined device can be seen in Fig. 7. Two fibers are mounted in a fiber array and are moved laterally, together with a fixed pitch, relative to underlying multi-mode grating couplers. The two branches of the interferometer are routed between the two multi-mode grating couplers in such a way that the resulting phase shifts in either MMGC add up rather than cancelling each other out [13]. Note that this requires one waveguide to be routed around the first MMGC as indicated by a red arrow.

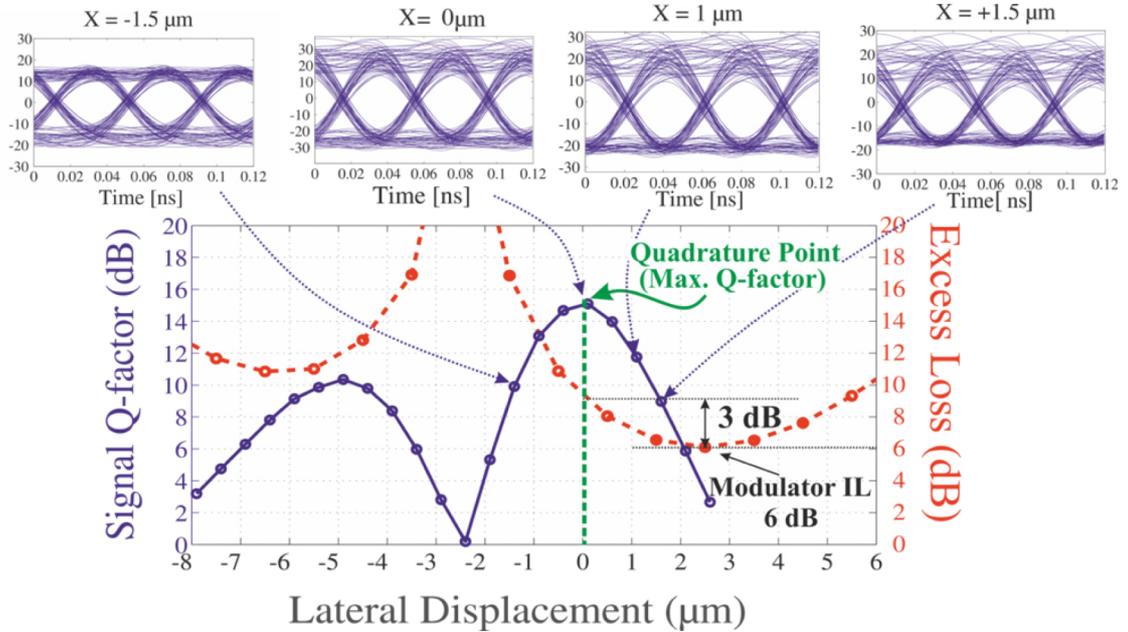

Fig. 8. Recorded signal Q-factors and extinctions with an on-resonance optical carrier as a function of the transverse fiber array displacement. The maximum Q-factor / OMA are obtained for a fiber positioning corresponding to the quadrature point, at which the extinction is 3 dB above its minimum. Recorded eye diagrams are shown for selected fiber array positions.

The temperature tolerant MZM itself corresponds to exactly the same design as reported in the previous section, and has thus the same temperature and wavelength tolerance. Fig. 8 shows the signal Q-factor recorded at the output of the device, as well as the extinction of the in-coupled light occurring inside the MZM (grating coupler losses, as measured at the optimal lateral displacement, normalized out), as a function of the transverse fiber array displacement. The Q-factor was measured with a 13 dBm laser power and a Finisar/U2T XPRV2021A 40 GHz bandwidth photoreceiver with an input referred transimpedance amplifier noise density specified to be below 40 $pA/\sqrt{Hz}$ and a responsivity between 0.5 A/W and 0.75 A/W. It can be seen that the signal Q-factor is maximized at a fiber displacement at which the on-resonance optical carrier excess loss is exactly 3 dB above its minimum, i.e., at a transverse displacement corresponding to the interferometer being biased at its quadrature point. Moreover, it can be seen that the sensitivity of the modulator bias point on fiber misalignment is approximately doubled relative to Fig. 2, by nature of two MMGCs being utilized in the device together with a fiber array and the incurred displacement induced phase shift thus being doubled. The signal Q-factor, and thus the OMA, remain within 3 dB of their maxima in a transverse fiber misalignment range of ±1.8 μm.

## V. Conclusions

In conclusion, we have shown a resonantly enhanced Mach-Zehnder modulator driven as a lumped element load with a ~20X reduction of power consumption compared to a linear travelling wave device with identical output optical modulation amplitude. The optical modulation amplitude is maintained within 3 dB of its maximum in a 55°C temperature range, which can be further increased by increasing the number of collectively driven RRMs on each branch. The Mach-Zehnder modulator was passively biased at its quadrature point by selective placement of the input and output fibers on top of multi-mode grating couplers also used as the input and output splitters of the interferometer.

## Acknowledgment

The authors gratefully acknowledge funding from the European Research Council under grant agreement 279770 and from the Excellence Initiative of the German State and Federal governments.